\documentclass[showpacs, pra,onecolumn,preprintnumbers ,amsmath, amssymb, superscriptaddress, aps]{revtex4-2}
\usepackage{color}
\usepackage{amsmath,amssymb}
\usepackage{pifont}
\usepackage{amssymb}  
\usepackage{bbold}
\usepackage{float}
\usepackage{subfloat}

\usepackage[caption=false]{subfig}
\usepackage{tikz}
\usepackage{makecell}
\usepackage{subfig}
\usepackage{pifont}   
\usepackage{graphicx} 
\graphicspath{{Figures/}}
\usepackage{dcolumn}  
\usepackage{bm}       
\usepackage{multirow} 
\usepackage{placeins}
\usepackage[colorlinks]{hyperref}
\hypersetup{colorlinks=true,linkcolor=blue}
\usepackage{mathtools}
\usepackage{appendix}

\newcommand{\bb}[1]{{\color{blue}#1}}

\def \be{\begin{align}}
	\def \ee{\end{align}}
\def \bea{\begin{eqnarray}}
	\def \eea{\end{eqnarray}}

\begin{document}
	
	\title{Effect of laser field and magnetic flux on scattering in graphene quantum dots}

	\author{Mohammed El Azar}
	\affiliation{Laboratory of Theoretical Physics, Faculty of Sciences, Choua\"ib Doukkali University, PO Box 20, 24000 El Jadida, Morocco}
	
		\author{Ahmed Bouhlal}
	\affiliation{Laboratory of Theoretical Physics, Faculty of Sciences, Choua\"ib Doukkali University, PO Box 20, 24000 El Jadida, Morocco}

	\author{Hocine Bahlouli}
	\affiliation{Physics Department and IRC Advanced Materials, King Fahd University of Petroleum and Minerals, Dhahran 31261, Saudi Arabia}
	
	\author{Ahmed Jellal}
	\email{a.jellal@ucd.ac.ma}
	\affiliation{Laboratory of Theoretical Physics, Faculty of Sciences, Choua\"ib Doukkali University, PO Box 20, 24000 El Jadida, Morocco}

	\begin{abstract}

		We show how Dirac electrons interact with a graphene quantum dots (GQDs) when exposed to both a magnetic flux and circularly polarized light. After obtaining the solutions of the energy spectrum, we compute the scattering coefficients. These allow us to show how efficiently the electrons diffuse and how their probability density is distributed in space. Our results show that light polarization is key in controlling electron scattering. It affects electron localization near the GQDs and the strength of the scattering coefficients. We also investigate how light intensity and magnetic flux affect the formation of quasi-bound states. In addition, the electrostatic potential reduces the density of scattering states and fine-tunes the interaction between electrons and the quantum dot. This research improves our understanding of electron behavior in graphene nanostructures and suggests new ways to control electronic states at the quantum level.

	\end{abstract}

	\pacs{}
	\maketitle

	\section{Introduction}

Graphene is a fascinating two-dimensional material composed of carbon atoms arranged in a hexagonal lattice. Its unique structure leads to impressive electronic properties, especially at low energies. Under these circumstances, charge carriers behave like massless Dirac fermions \cite{Novoselov05,Katsnelson07,CastroNeto09,Castro10}, illustrating physical phenomena that were available only in relativistic particle physics. This relationship has attracted considerable interest, linking condensed matter physics with quantum field theory. Ongoing research on graphene is revealing valuable insights that benefit both materials science and fundamental physics.
Two-dimensional nanomaterials, in particular graphene, continue to attract considerable scientific and technological interest due to their outstanding electronic, optical, and mechanical properties. Since its first isolation in 2004 \cite{Novoselov04}, graphene has become a benchmark material for the study of fundamental quantum phenomena and for the development of new applications, in particular in optoelectronics as sensors, biosensors and in communication technologies \cite{novoselov2007electronic,bonaccorso2010graphene,geim2007rise,konstantatos2012hybrid}.
Among the most intriguing properties of graphene, electron scattering, and in particular,  electron trapping and confinement at the nanoscale, occupy a prominent place in research. The band structure of graphene, characterized by its Dirac cones, gives rise to relativistic electron transport \cite{pereira2009strain, yang2018structure}, where Klein tunneling prevents the static localization of electrons in confined regions \cite{Katsnelson06klein}. However, recent studies have highlighted the possibility of forming quasi-bound states, where electrons are temporarily trapped, in graphene quantum dots (GQDs) through the use of various techniques ranging from the application of magnetic fields \cite{azar24, Pena22driven, azar2024effects, weymann2015transport} to substrate engineering \cite{gutierrez2016klein, lee2016synergetic} via the use of light \cite{Penalight}.
The tuning of these quasi-bound states and the control of their lifetime are of paramount importance both from the fundamental physics perspective and technological applications. In the field of fundamental physics, the {controlled} localization of electrons is required for the development of quantum devices such as single electron transistors \cite{xu2022atomically, gao2024visualization}, memories, or in the design of qubits \cite{trauzettel2007spin}. In the field of technological applications, the control of these quasi-bound states is important for the development of quantum devices. Among the most promising control methods, the interaction of electrons with polarized light has attracted particular interest \cite{farias2013photoexcitation, scholz2014absorption, el2024transport}. The use of polarized light, such as circularly polarized light, has emerged as a novel approach to manipulate electron dynamics in graphene, allowing the control of the lifetime of quasi-bound states \cite{Penalight, Benlakhouy22}.
	
The use of graphene in electronic applications is hampered by the absence of a bandgap in its free standing graphene crystallite state. Several approaches have been developed to overcome this limitation, with nanostructuring of graphene showing particular promises. The first strategy involves shaping graphene into an effective one-dimensional structure known as nanoribbons \cite{Han2007}, through which it is possible to induce the appearance of a bandgap by confinement effects. This approach can be further enhanced using graphene quantum dots (GQDs), a zero-dimensional structure {which} by its very nature enhances quantum confinement \cite{Guttinger2009, Libisch10}. These GQDs exhibit remarkable properties reminiscent of those of isolated atoms {that} is why they are called "artificial atoms". 
The interest in these nanostructures extends to many areas of application. In the field of quantum computing, GQDs offer intriguing possibilities for information storage and data manipulation \cite{Li2013}. Their potential usage also extends to electronic devices, gas sensors, biosensors and photovoltaic and solar cell technologies \cite{Sun2013, Bacon2014}. However, a major challenge remains: the Klein tunneling effect \cite{Martino2007, Myoung2009, Esmailpour2018}, a unique feature of Dirac fermions in graphene, compromises the efficient confinement of electrons in these nanostructures. To overcome this obstacle, several scientific studies are actively exploring various solutions, including the use of magnetic fields to create magnetic confinement in GQDs \cite{Matulis09, Wang09, Ramezani09, Giavaras09, Heinisch2013, Guclu2013, Myoung2019, Pan2020, Belokda23, ElAzarflux24, ElazarBoosting24}. This approach opens new perspectives for the development of functional graphene based devices.

In this study, a comprehensive theoretical investigation of electron scattering in a graphene quantum dot under the influence of a magnetic flux and circularly polarized light irradiation is presented, with special emphasis on the nature of quasi-bound states. Using a numerical approach, we systematically investigate the effect of external control parameters, including light intensity and polarization, magnetic flux, and electrostatic potential, on electron scattering. The results of the study underscore the key role of light polarization in {controling the interaction} between electrons and the GQD. Furthermore, we show that external parameters can be used to modulate the lifetime of quasi-bound states, in particular,  by {varying} the intensity of light and illustrate how the effects of magnetic flux combine with those of light to provide a comprehensive understanding of the phenomenon. The results obtained contribute to the fundamental understanding of the quantum diffusion of electrons in graphene and open interesting avenues for the development of optoelectronic sensors and quantum devices based on this material.
	
The present work is structured to provide a comprehensive and systematic analysis of the behavior of Dirac electrons in a graphene quantum dot subjected to a magnetic flux and circularly polarized light irradiation. In Sec. \ref{sec1}, the theoretical framework necessary to understand the system is developed. This section is divided into three complementary parts: the first deals with the eigenspinors in the two valleys inside the quantum dot, while the second focuses on their behavior outside the dot. The third part is devoted to the phenomenon of diffusion, establishing the essential theoretical tools for characterizing the behavior of electrons in this complex system. Sec. \ref{res} presents the quantitative results of the study, providing an in-depth understanding of the electron scattering phenomena and the influence of the various system parameters. Finally, Sec. \ref{conc} summarizes the main results of the study and highlights their implications for the physics of graphene nanostructures.
	
\section{Mathematical formalism}\label{sec1}

\begin{figure}[h]
		\centering 
		\includegraphics[scale=0.8]{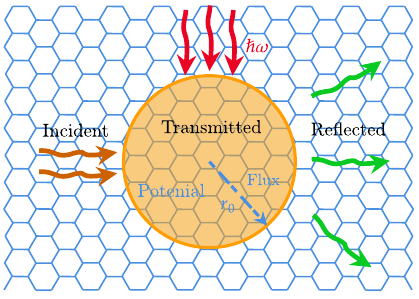}	
		\caption{Schematic profile of a graphene quantum dot (GQD) of radius $r_0$.		
		}\label{figsystem}
	\end{figure}

To illustration the physical picture of our system, consider the schematic profile shown in Fig. \ref{figsystem}. It shows a graphene quantum dot (GQD), represented as a yellow circle with radius $r_0$, placed in a two-dimensional plane. The GQD is exposed to a magnetic flux $\Phi$ and a beam of circularly polarized light. The light has a frequency $\omega$ and a polarization $\mathcal{P}$ and is directed perpendicular to the GQD along the $z$-axis. An incoming electron, indicated by the orange arrows, approaches the GQD and scatters. This scattering process produces a reflected wave outside the GQD and a transmitted wave inside the GQD.

{The motion of charge carriers in graphene quantum dots (GQDs) of radius $r_0$ subjected to laser field, magnetic flux and electrostatic potential can be described by the following Hamiltonian near Dirac point $K$
	\begin{equation}\label{ham1}
		H=  v_F\ \vec \sigma\cdot (\vec p + e \vec A(r) + \vec A(t)) + U\mathbb{I}_2
	\end{equation}
where the vector potential associated {with} the magnetic flux $\Phi$ is given by \cite{Heinl2013}
	\begin{equation}  \vec A(r)=\frac{h}{e}\frac{\Phi}{2\pi r}\hat{e}_{\theta}
	\end{equation}
and $\hat{e_{\theta}}$ is the unit vector along the azimuthal direction,  $U$ is the applied potential and $\sigma_i$ are Pauli matrices.
To avoid coupling of the two valleys $K$ and $K'$, we assume that the spatial extension of the flux line is large compared to the lattice constant. The vector potential of the light in plane wave representation in the case of an arbitrary polarization reads 
	\begin{align}
		{\boldsymbol{A}}(t)=A_0\left\{\left[\cos(\omega t)+{\mathcal{P}}\cos(\omega 
		t+\theta)\right]\vec e_x+\left[-\sin(\omega t)+{\mathcal{P}}\sin(\omega 
		t+\theta)\right]\vec e_y\right\}
	\end{align}
where $A_0$ is a real constant amplitude, $\omega$ the light frequency, $\mathcal{P}$ is the polarization, $\vec e_x$ and $\vec e_y$ 
are the unit vectors along the $x-$ and $y-$axis, respectively. Generally speaking, $\boldsymbol{A}(t)$ describes an elliptical state of 
polarization. After some algebra, we can show that \eqref{ham1}  can be transformed into
	\begin{equation}\label{Hamilt2}
		H =v_F \vec \sigma \cdot (\vec p +e \vec A(r) ) + \alpha \sigma_z 
		+ U\mathbb{I}_2
	\end{equation}	
where we have set the energy gap $\alpha=- \frac{(ev_FA_0)^2}{\hbar\omega}({\mathcal{P}}^2-1)$, which is generated  by transforming \eqref{ham1} into the  time independent Hamiltonian \eqref{Hamilt2}. More details about such transformation can be found in literature \cite{Penalight,Benlakhouy22}.

To determine the fermion eigenspinors in monolayer graphene subjected to magnetic flux, we consider the basis
	$\Psi(r ,\theta)=
	(\Psi_A, \Psi_B)^T$ 
in polar coordinates $(r,\theta)$ due to system symmetry. 
Owing to the fact that $[H, J_z]=0$, then the separability imposes 
	\begin{equation}
		\Psi(r ,\theta)=e^{im\theta}\begin{pmatrix}\chi_A(r) \\ ie^{i \theta}\chi_B(r)  \end{pmatrix} 
	\end{equation}
where $m = 0,\pm1,\pm2,\cdots $ are eigenvalues of  
$L_z$ associated {with} the total angular momentum $J_z= L_z+\frac{\hbar}{2}\sigma_z$. To proceed further, let us 
write the Hamiltonian \eqref{ham1} 
	\begin{equation}
		H= \hbar v_F\begin{pmatrix}
			u_+ & 
			\pi_+
			\\ \pi_- &  u_-
		\end{pmatrix}
	\end{equation}
in terms of the operators 
	\begin{align}
		\pi_{\pm }=e^{\mp i { \theta}}\left(-i\frac{\partial}{\partial \xi}\mp\frac{1}{\xi}\frac{\partial}{\partial \theta}\mp i\frac{\Phi}{\xi}\right)
	\end{align}
where we have introduced the change of variable $\xi=\frac{r}{r_0}$,    
and the  dimensionless quantities  $\varepsilon=\frac{Er_0}{\hbar v_F}$,  $u =\frac{U r_0}{\hbar v_F}$, and $\delta=\frac{\alpha r_0}{\hbar v_F}$, with $u_\pm= u\pm \delta$. The flux $\Phi$ in all over the manuscript from now on is measured in units of quantum flux $\Phi_0=\frac{\hbar}{e}$.
In the next step, we proceed by solving the eigenvalue equation $
H\Psi(\xi,\theta)=\varepsilon \Psi(\xi,\theta)$. Indeed, inside GQDs ($r<r_0$) we find two coupled equations
	\begin{align}
		&
		\left(\frac{\partial}{\partial \xi}+\frac{ m+1 +\Phi}{\xi} 
		\right)\chi_B=\left(\varepsilon-u_+\right)\chi_A\label{R-b}\\
		&
		\left(\frac{\partial}{\partial \xi}-\frac{ m+ \Phi}{\xi}
		\right)\chi_A=-\left(\varepsilon-u_-\right)\chi_B \label{R+b}.
	\end{align}
By injecting \eqref{R+b} into \eqref{R-b}, we obtain a second order differential equation for $\chi_A$
	\begin{align}
		\left[\frac{\partial^2}{\partial \xi^2}+\frac{1}{\xi}\frac{\partial}{\partial \xi}-\frac{1}{\xi^2}\left( m+\Phi\right)^2+ (\varepsilon-u)^2-\delta^2\right]\chi_A=0
	\end{align}
and the corresponding {regular} solutions are the Bessel functions
	\begin{eqnarray}\label{chiA}
		\chi_A(\xi)\sim J_{\nu}(\lambda\xi)
	\end{eqnarray}
where we set $\lambda=\sqrt{|(\varepsilon - u)^2- \delta^2}$ and a new quantum number $\nu = m+\Phi$ depending on the flux $\Phi$. If we now replace \eqref{chiA} by \eqref{R+b} and
use the recurrence relations of  Bessel functions
\begin{align}
&
J_n^{\prime}(x)=\frac{n}{x}J_{n}(x)-J_{n+1}(x)
\end{align}
we {obtain} the second component 
	\begin{align}
\chi_B(\xi)=\mu J_{\nu+1}(\lambda\xi)
\end{align}
 and we have set $\mu = \frac{\lambda}{\varepsilon-u_-}$ with the condition $\varepsilon\neq u_-$. 
Finally, the eigenspinors inside the GQD {read} 
	\begin{equation}
		\Psi(\xi ,\theta)=e^{im\theta}\begin{pmatrix}J_{\nu}(\lambda\xi) \\ i \mu e^{i \theta}   J_{\nu+1}(\lambda\xi)  \end{pmatrix} .
	\end{equation}
These results are different from the energy spectrum of a circular graphene quantum dot subjected to a perpendicular magnetic field \cite{Schnez2008}. Based on the solutions of the Dirac equation derived above, the transmitted wave is expressed as follows		
		\begin{equation}\label{sp1}
			\psi_t(\xi ,\theta)=\sum_{m=-\infty}^{\infty} i^ma_m^{t}e^{im\theta}\begin{pmatrix}J_{\nu}(\lambda\xi) \\ i \mu e^{i \theta}  J_{\nu+1}(\lambda\xi)  \end{pmatrix}. 
		\end{equation}

To study electron diffusion through a quantum dot in graphene, we first consider an electron approaching the GQDs under normal incidence. The energy of the incident electron is given by $E = \hbar v_F q$, where $q$ is the associated wave number. After interaction, the electron can either be reflected or transmitted through the dot. This process is shown schematically in Fig. \ref{figsystem}. The incident and reflected waves involved in this process can be expressed by the following equations
	\begin{align}
		&\label{sp2}
		\psi_i(\xi, \theta)=
        e^{i q \xi \cos \theta}\binom{1}{1}= \sum_{m=-\infty}^{\infty} i^m e^{i m \theta}\binom{J_m(q \xi) }{ie^{i\theta} J_{m+1}(q \xi) }
\\
&\label{sp3}
	\psi_r(\xi, \theta)=
    \sum_{m=-\infty}^{\infty} i^m b_m^{r} e^{i m \theta}\binom{H_m(q \xi) }{i e^{i \theta} H_{m+1}(q \xi)}
\end{align}
where  \eqref{sp2} represents a plane wave and is expressed as an infinite sum of well-defined orbital angular momentum states using the Jacobi-Anger expansion $J_m (x)$ (Bessel function of the first kind). While  \eqref{sp3} illustrates the decomposition of the reflected wave into partial waves, where $H_m (x)$ represents the Hankel functions of the first kind of order $m$ \cite{Cserti2007}.

\subsection{Scattering problem}
To analyze the scattering problem, we first determine the scattering coefficients $a_m^{t}$ and $b_m^{r}$. This can be done using the boundary condition at the interface $\xi = 1$  $(r = r_0)$. Then {from}  \eqref{sp1}, \eqref{sp2} and \eqref{sp3}, we can write the continuity equation
\begin{equation}
\psi_i(1, \theta) + 	\psi_r(1, \theta) = \psi_t(1, \theta) 
\end{equation}
After some algebras, we get 
\begin{align}
&a_m^{t}(\lambda)=\frac{J_{m}(q)H_{m+1}(q) - J_{m+1}(q)H_{m}(q) }{J_{\nu}(\lambda)H_{m+1}(q) -\mu J_{\nu+1}(\lambda)H_m(q)}\\
&b_m^{r}(\lambda)=\frac{\mu J_{\nu+1}(\lambda)J_m(q) - J_{\nu}(\lambda)J_{m+1}(q) }{J_{\nu}(\lambda)H_{m+1}(q) -\mu J_{\nu+1}(\lambda)H_m(q)} \label{31b}
\end{align}
with the quantum number is $\nu=m+\Phi$.

To provide a more quantitative description of the scattering, we will introduce the concept of scattering efficiency \( Q \).
To enhance our study of the scattering problem of Dirac fermions in GQDs of different sizes, we study \( Q \). This is defined as the ratio of the scattering cross section to the geometrical cross-section \cite{azar24, el2024transport}
\begin{equation}
	Q=\frac{4}{q r_0} \sum_{m=-\infty}^{\infty}\left|b_m^{r}(\lambda)\right|^2.
\end{equation}
Recall that the coefficients \( b_m^r \) in \eqref{31b} depend on the incident energy, the energy gap \( \delta \), the magnetic flux \(\Phi \), the laser amplitude \( A_0 \) and the polarization \( {\cal P} \). This dependence provides several control parameters for our physical system, allowing us to explore and discuss different aspects of the scattering phenomenon occurring in our system.

\section{NUMERICAL ANALYSIS}\label{res}

In the present study, a comprehensive investigation of electron scattering in graphene quantum dots (GQDs), including the interplay of magnetic flux and laser irradiation, is performed. The scattering efficiency $Q$, which quantifies the ability of the GQD to scatter incident electrons, is the main parameter used to characterize the scattering properties of the system. Our study systematically analyzes the influence of several key factors: quantum dot size $r_0$, magnetic flux $\Phi$, electrostatic potential $U$, light intensity $I_L$, and light polarization state $\mathcal{P}$. Laser irradiation is characterized by its intensity, defined as $I_L$, which is directly proportional to the square of the amplitude of the vector potential of the electromagnetic wave. More precisely, $I_L$ can be expressed as $\epsilon_0\omega^2A_0^2$, where $\epsilon_0$ is the permittivity of the vacuum and $\omega$ is the angular frequency of the light wave. The quantity $I_L$ is used as a control parameter to assess the effect of laser light on electron transport and confinement in GQDs. The present study aims to elucidate how light influences electron confinement. In our {computational work}, the frequency of the light is kept constant at $5 \times 10^{14} \text{ s}^{-1}$.
This value is in the low energy regime, where the unique properties of graphene are most pronounced. Our numerical calculations have revealed a variety of scattering phenomena, including distinct resonances and complex correlations between physical parameters. A particular focus of our study is the investigation of the different excited scattering modes and their effect on the scattering efficiency $Q$.

Fig. \ref{fig2} illustrates the scattering efficiency $Q$ as a function of light intensity $I_L$ and light polarization $\mathcal{P}$ for an incident electron energy fixed  $E=20 \text{ meV}$ and a GQD radius  $r_0=70 \text{ nm}$. Fig. \ref{fig2} is subdivided into 12 panels, with each panel corresponding to a specific combination of electrostatic potential $U$ ($0$, $5 \text{ meV}$, and $10 \text{ meV}$) and magnetic flux $\Phi$ ($0$, $1/2$, $3/2$, and $5/2$). A general observation reveals the presence of complex scattering patterns, where the scattering efficiency $Q$ fluctuates non-monotonically as a function of $I_L$ and $\mathcal{P}$. These patterns, which reflect the ability of the quantum dot to scatter incident electrons, are sensitive to system parameters and reveal a complex interplay between light, electrostatic potential, and magnetic flux.
Figs. \ref{fig2}\bb{(a)-(d)} correspond to zero electrostatic potential ($U = 0 \text{ meV}$). Fig. \ref{fig2}\bb{(a)}, with $\Phi=0$, reveals regular and symmetrical resonance patterns, which are characteristic of isotropic scattering in the absence of magnetic flux. The introduction of a magnetic flux of magnitude $1/2$ in Fig. \ref{fig2}\bb{(b)} produces a discernible alteration in the patterns, manifested by the emergence of distinct periodic structures. Figs. \ref{fig2}\bb{(c,d)}, corresponding respectively to $\Phi = 3/2$ and $\Phi = 5/2$, show a progressive enrichment of the scattering patterns, which is indicative of the increasing influence of magnetic flux on quantum interference.
As illustrated in Figs. \ref{fig2}\bb{(e)-(h)}, where $U = 5 \text{ meV}$, a substantial modification of the scattering characteristics is evident. Fig. \ref{fig2}(\bb{e)} demonstrates that, in the absence of magnetic flux, the electrostatic potential significantly modifies the resonance patterns in comparison with Fig. \ref{fig2}(a). As the magnetic flux increases, Figs. \ref{fig2}\bb{(f)-(h)} demonstrate a general decline in the scattering efficiency $Q$, accompanied by a progressive deformation of the areas of strong scattering. This evolution underscores the intricate interplay between electrostatic potential and magnetic flux.
In Figs. \ref{fig2}\bb{(i)-(l)}, where $U = 10 \text{ meV}$, the most intricate patterns emerge. Fig. \ref{fig2}\bb{(i)}, in the absence of magnetic flux, already exhibits highly non-uniform scattering patterns, indicating the substantial impact of the high electrostatic potential. 
In Figs. \ref{fig2}\bb{(j)-(l)}, the gradual increase in magnetic flux gives rise to the emergence of highly non-linear scattering structures.  These panels reveal more localized resonance zones and increased sensitivity to variations in light intensity and polarization.
A comparative analysis of the twelve panels reveals several noteworthy trends. First, an increase in electrostatic potential (illustrated by vertical progression of \bb{(a)}$\rightarrow$\bb{(i)}, \bb{(b)}$\rightarrow$\bb{(j)}, \bb{(c)}$\rightarrow$\bb{(k)}, and \bb{(d)}$\rightarrow$\bb{(l)}) results in a systematic {display} of the scattering patterns. Second, the enhancement of magnetic flux (across each line) modulates the fine structure of the resonances, introducing characteristic asymmetries into the scattering patterns. This modulation becomes particularly pronounced for high values of electrostatic potential.

 \begin{figure}[ht]
 	\centering
 	\includegraphics[scale=0.33]{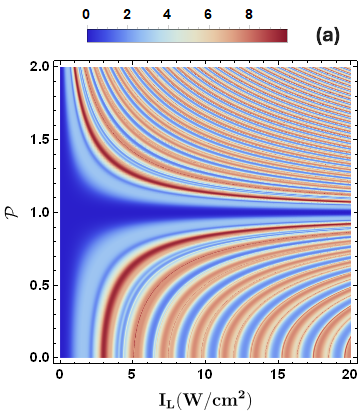}\includegraphics[scale=0.33]{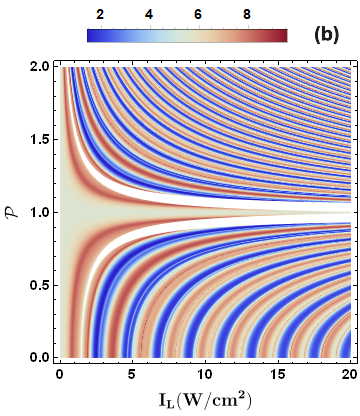}\includegraphics[scale=0.33]{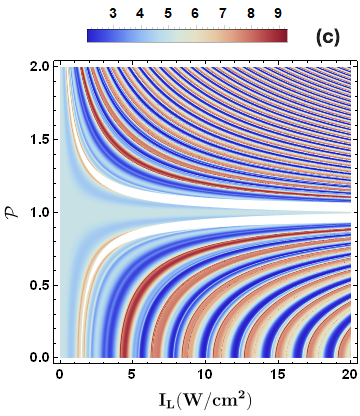}\includegraphics[scale=0.33]{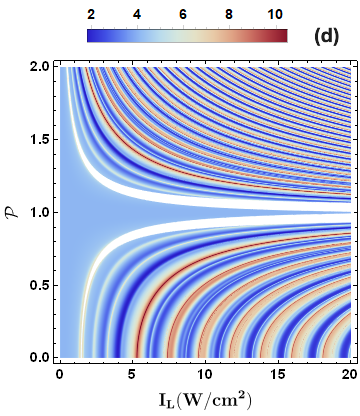}\\
 		\includegraphics[scale=0.33]{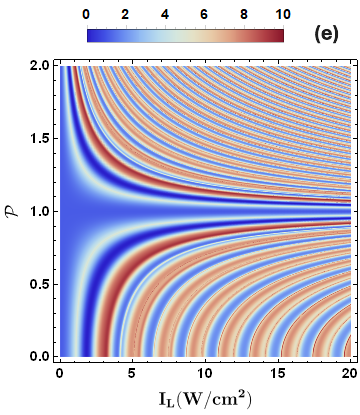}\includegraphics[scale=0.33]{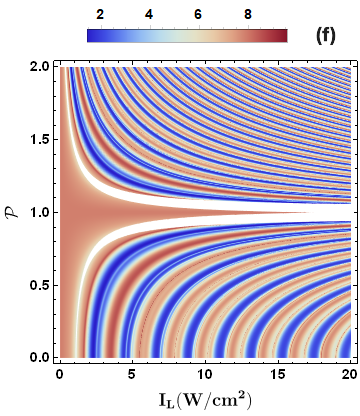}\includegraphics[scale=0.33]{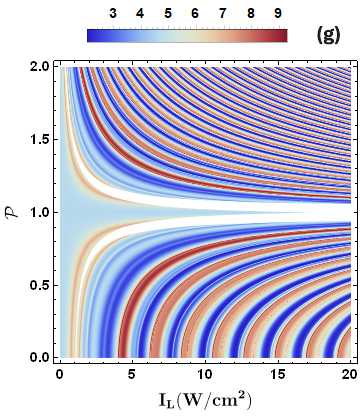}\includegraphics[scale=0.33]{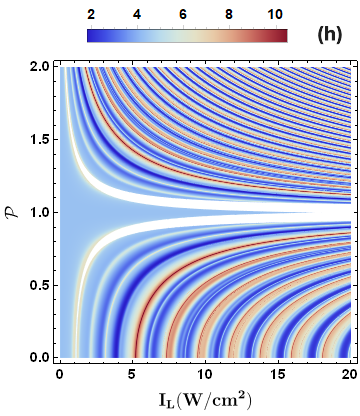}\\
 			\includegraphics[scale=0.33]{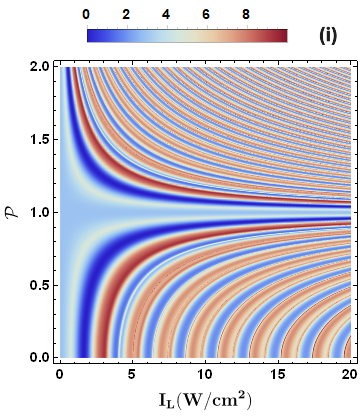}\includegraphics[scale=0.33]{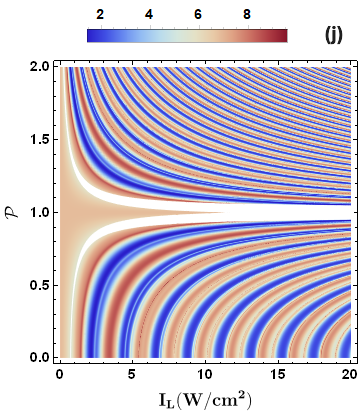}\includegraphics[scale=0.33]{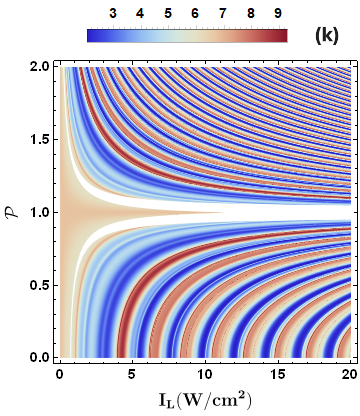}\includegraphics[scale=0.33]{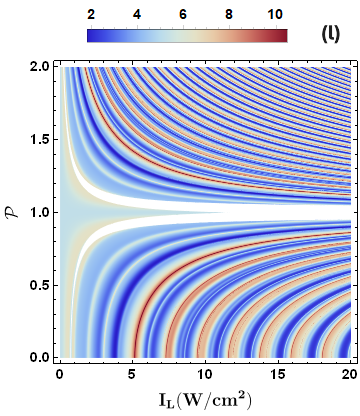}
 	\caption{ The scattering efficiency $Q$ as a function of the light intensity $I_L$ and light polarization $\mathcal{P}$ for $E = 20 \text{ meV}$ and $r_0 = 70 \text{ nm}$ by changing the value of electrostatic potential $U$: (a, b, c, d) for $U = 0 \text{ meV}$, (e, f, g, h) for $U = 5 \text{ meV}$ and (i, j, k, l) for $U = 10 \text{ meV}$. {The plots have been generated} by fixing the value of magnetic flux $\Phi$ such that (a, e, i) for $\Phi = 0$, (b, f, j) for $\Phi = 1/2$, (c, g, k) for $\Phi = 3/2$ and (d, h, l) for $\Phi = 5/2$.}
 	\label{fig2}
 \end{figure}

Fig. \ref{fig3} presents an in-depth analysis of the scattering efficiency $Q$ as a function of light polarization $\mathcal{P}$ for a GQD characterized by an incident energy  $E=20 \text{ meV}$ and a radius $r_0=70 \text{ nm}$. The study systematically compares the influence of different parameters in eight different configurations, revealing the relationship between electrostatic potential $U$, magnetic flux $\Phi$, and light intensity $I_L$.
Figs. \ref{fig3}\bb{(a)-(d)} correspond to $I_L = 3 \text{ W/cm}^2$, and they reveal the evolution of scattering for different values of magnetic flux $\Phi$. In Fig. \ref{fig3}\bb{(a)}, where $\Phi = 0$, the three curves (blue for $U = 0 \text{ meV}$, green for $U = 5 \text{meV}$, and red for $U = 10 \text{ meV}$) demonstrate distinct oscillations in scattering efficiency as a function of polarization. The blue curve demonstrates the most uniform oscillations, indicating that the absence of electrostatic potential promotes a more coherent system response. In contrast, the green and red curves exhibit increasingly intricate behavior, suggesting that an increase in electrostatic potential leads to substantial alterations in diffusion mechanisms.
As shown in Figs. \ref{fig3}\bb{(b)-(d)}, the introduction of increasing magnetic flux (denoted as $\Phi = 1/2$, $3/2$, and $5/2$, respectively) leads to significant changes in the scattering profiles. The oscillation patterns undergo an evolution, accompanied by the emergence of more complex structures and changes in the positions of maxima and minima. This evolution is indicative of the influence of the magnetic flux on the quantum interference within the system.
Figs. \ref{fig3}\bb{(e)-(h)}, corresponding to a higher light intensity ($I_L = 5 \text{ W/cm}^2$), show qualitatively different characteristics. It has been found that increasing the light intensity generally increases the scattering efficiency and modifies the fine structure of the oscillations. In Fig. \ref{fig3}\bb{(e)}, with zero magnetic flux ($\Phi = 0$), all three curves display larger oscillation amplitudes compared to those in Fig. \ref{fig3}\bb{(a)}, indicating a stronger coupling between light and electrons. Figs. \ref{fig3}\bb{(f)-(h)} illustrate how this enhancement responds to increasing magnetic flux, resulting in more complex scattering patterns.
A notable feature is the systematic effect of increasing electrostatic potential (from blue to red curve) on the scattering profiles across all panels. The progressive shift of the maxima and the modification of the oscillation amplitudes suggest that the electrostatic potential plays a key role in modulating the available scattering states.
A comparison of the two lines (Figs. \ref{fig3}\bb{(a)-(d)} and Figs. \ref{fig3}\bb{(e)-(h)}) reveals the non-trivial effect of light intensity on the scattering properties. Increasing the incident light intensity $I_L$ from $3$ to $5 \text{ W/cm}^2$ does not simply result in a uniform amplification of the scattering patterns. Instead, it leads to a complex rearrangement of the oscillation profiles, which is particularly evident in configurations with non-zero magnetic flux.
  
   \begin{figure}[ht]
  	\centering
  	\includegraphics[scale=0.34]{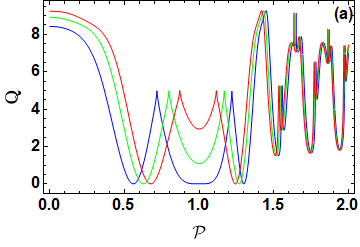}\includegraphics[scale=0.34]{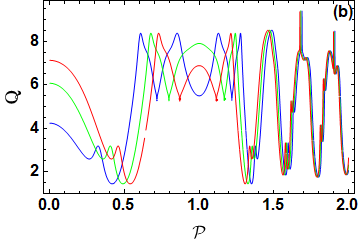}\includegraphics[scale=0.34]{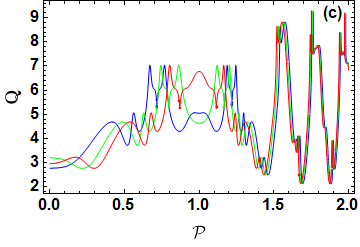}\includegraphics[scale=0.34]{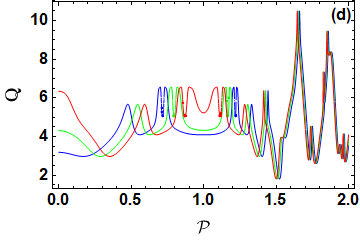}\\
  	\includegraphics[scale=0.34]{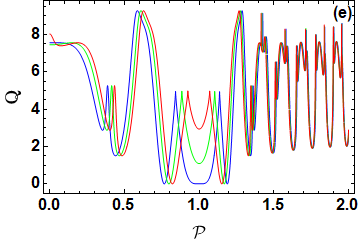}\includegraphics[scale=0.34]{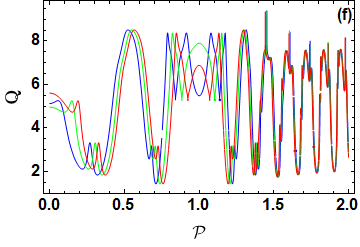}\includegraphics[scale=0.34]{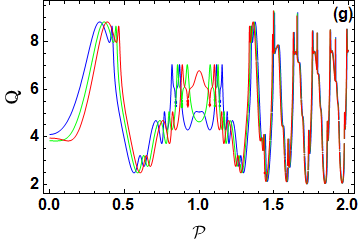}\includegraphics[scale=0.34]{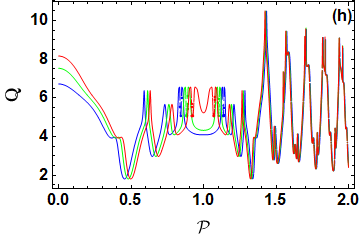}
  	\caption{ The scattering efficiency $Q$ as a function of the light polarization $\mathcal{P}$ for $E = 20 \text{ meV}$ and $r_0 = 70 \text{ nm}$ by changing the values of the value of electrostatic potential $U$ such that blue line for $U = 0 \text{ meV}$, green line for $U = 5 \text{ meV}$ and red line for $U = 10 \text{ meV}$. In the ours plots, we {vary the} magnetic flux $\Phi$ such that (a, e) for $\Phi = 0$, (b, f) for $\Phi = 1/2$, (c, g) for $\Phi = 3/2$ and (d, h) for $\Phi = 5/2$. {The generated plots were obtained} by fixing the value of light intensity $I_L$ such that (a, b, c, d) for  $I_L = 3 \text{ W/cm}^2$ and (e, f, g, h) for  $I_L = 5 \text{ W/cm}^2$.}
  	\label{fig3}
  \end{figure}
As shown in Fig. \ref{fig3a}, a comprehensive analysis of the scattering efficiency $Q$ as a function of light polarization $\mathcal{P}$ is performed for incident energy $E= 20 \text{ meV}$ and quantum dot radius $r_0=70 \text{ nm}$. This study is carried out by considering different values of magnetic flux $\Phi$ and electrostatic potential $U$. Figs. \ref{fig3a}\bb{(a)-(d)} correspond to zero electrostatic potential $U$, while Figs. \ref{fig3a}\bb{(e)-(h)} correspond to a potential $U$ of $10 \text{ meV}$. In each panel, the orange curve represents the total scattering efficiency, while the colored curves highlight the contribution of each scattering mode ($m=-1$ in brown, $m=0$ in blue, $m=1$ in green, $m=2$ in red, $m=3$ in magenta, and $m=4$ in violet).
First, an analysis of the total scattering efficiency (orange curve) reveals oscillations as a function of $\mathcal{P}$, with a primary peak observed in most panels at $\mathcal{P}=0$. However, the amplitude of these peaks varies depending on the electrostatic potential $U$. Figs. \ref{fig3a}\bb{(a)–(d)} ($U = 0 \text{ meV}$) demonstrate more pronounced peaks and an overall higher amplitude, while Figs. \ref{fig3a}\bb{(e)–(h)} ($U = 10 \text{ meV}$) show weaker and shifted peaks. It is also evident that the shape of the peaks is modified by the value of $U$; the peaks are flatter for $U=0$ and sharper for $U=10 \text{ meV}$. These variations suggest that the electrostatic potential influences not only the intensity of the diffusion, but also the underlying resonance mechanism.
A subsequent examination of the colored curves then reveals the characteristics of each scattering mode. For the $m=-1$ and $m=0$ modes, the amplitudes are smaller than for the $m=1$ mode and show similar oscillatory behavior, while the $m=2$ mode shows oscillatory behavior, but its shape and amplitude are very different from those of the $m=-1$ and $m=0$ modes. The $m=3$ and $m=4$ modes show different behavior: the $m=3$ mode has a larger amplitude and less pronounced oscillations, while the $m=4$ mode is almost absent. It is important to note that, as can be seen in Figs. \ref{fig3a}\bb{(a)} to \bb{(h)}, each mode is present at all values of flux and potential, but their contribution varies considerably {depending on} these parameters.
Furthermore, it can be seen that the peaks associated with each mode are located in regions where the total scattering efficiency is large, showing a direct link between the scattering resonances and the contributions of specific modes. It is also interesting to note that for the same potential, increasing the flux increases the amplitude and number of peaks of the $m=-1$ and $m=0$ modes, and generally shifts the resonances to higher values of $\mathcal{P}$. As noted above, the amplitude of the total diffusion efficiency is reduced for $U = 10 \text{ meV}$ and the resonances are more pronounced for the lowest potentials. Furthermore, a comparison of the values of total $Q$ for different fluxes shows that the {height} of the peaks generally decreases with increasing flux for $U=0$ and increases for $U = 10 \text{ meV}$. This shows that there is a complex interplay between electrostatic potential and magnetic flux in {controlling} electron diffusion. In conclusion, Fig. \ref{fig3a} presents a comprehensive study of the electron diffusion mechanism in GQDs. It is clear from the figure that the scattering efficiency {depends} on a multifaceted interaction between the polarization of the light, the magnetic flux, the electrostatic potential, and the various scattering modes.
   \begin{figure}[ht]
	\centering
	\includegraphics[scale=0.34]{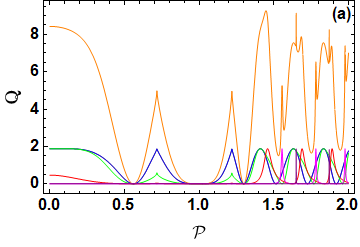}\includegraphics[scale=0.34]{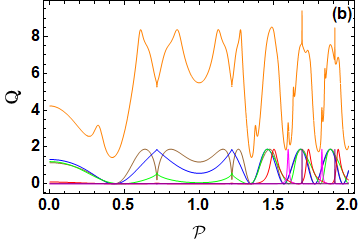}\includegraphics[scale=0.34]{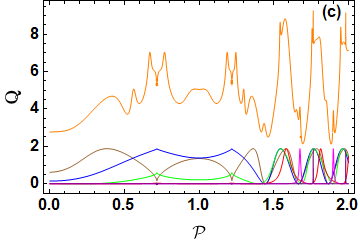}\includegraphics[scale=0.34]{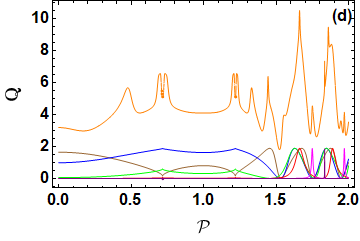}\\
	\includegraphics[scale=0.34]{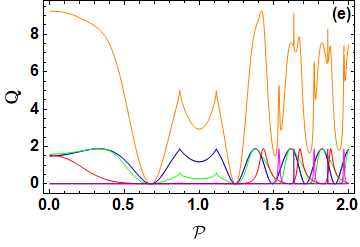}\includegraphics[scale=0.34]{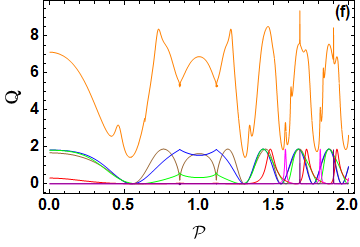}\includegraphics[scale=0.34]{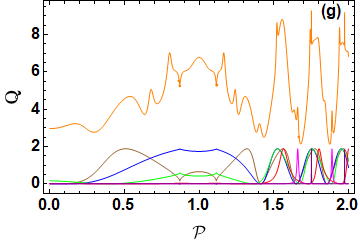}\includegraphics[scale=0.34]{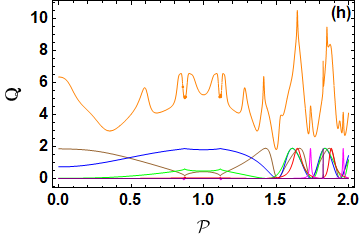}
	\caption{ The scattering efficiency $Q$ is plotted as a function of light polarization $\mathcal{P}$ for $E = 20 \text{ meV}$, $I_L = 3 \text{ W/cm}^2$, and $r_0 = 70 \text{ nm}$. The plots are generated by varying the electrostatic potential $U$ and the magnetic flux $\Phi$. (a, b, c, d) for $U = 0 \text{ meV}$, while (e, f, g, h) for $U = 10 \text{ meV}$. Within each panel, the orange line represents the total scattering efficiency $Q$. The colored lines represent the contribution of individual scattering modes, labeled by the angular momentum quantum number $m$: brown line for $m = -1$, blue line for $m = 0$, green line for $m = 1$, red line for $m = 2$, magenta line for $m = 3$, and purple line for $m = 4$. The magnetic flux $\Phi$ is varied as follows: (a, e) for $\Phi = 0$, (b, f) for $\Phi = 1/2$, (c, g) for $\Phi = 3/2$, and (d, h) for $\Phi = 5/2$.}
	\label{fig3a}
\end{figure}

As shown in Fig. \ref{fig4}, the analysis examines the influence of the quantum dot radius $r_0$, the electrostatic potential $U$, the magnetic flux $\Phi$, and the light intensity $I_L$ on the electron scattering efficiency $Q$. The analysis shows several significant trends. First, increasing the electrostatic potential $U$, as illustrated in the series of plots from top to bottom, results in a shift of the diffusion efficiency peaks to higher values of $r_0$ and a reduction in their amplitude. This behavior can be interpreted as an increased difficulty for electrons to penetrate the quantum dot as the electrostatic potential $U$ increases, which could be related to the results of previous studies on the effect of light \cite{Penalight}, where we see how the presence of light facilitates electron penetration. Furthermore, the reduced amplitude of the peaks suggests a modification of the scattering mechanism itself under the influence of the electrostatic potential.
Subsequently, a varying magnetic flux $\Phi$ causes variations in the structure of the $Q$ peaks. For example, a comparison of Figs. \ref{fig4}\bb{(a)-(d)} shows that the peaks {become} broader for a flux of $\Phi=0$, thinner and of higher amplitude for a flux of $\Phi=1/2$. These observations suggest that the magnetic flux influences the localization of resonances and their intensity, which could be related to an interplay between magnetic confinement and the energy of quasi-bound states. Furthermore, a comparison of Figs. \ref{fig4}\bb{(a)-(d)} and Figs. \ref{fig4}\bb{(e)-(h)} shows that the light intensity has a significant effect on the scattering resonances. The peaks become more pronounced, narrower and shifted to higher $r_0$ values as $I_L$ increases, suggesting that light enhances electron trapping. This hypothesis is further supported by the observation that scattering is significantly reduced at lower intensities.
   \begin{figure}[ht]
	\centering
	\includegraphics[scale=0.34]{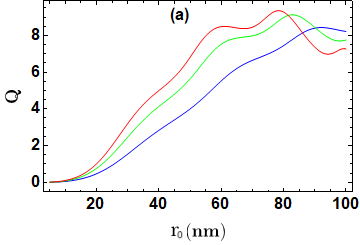}\includegraphics[scale=0.34]{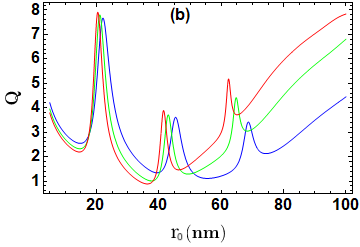}\includegraphics[scale=0.34]{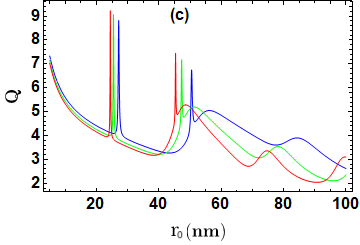}\includegraphics[scale=0.34]{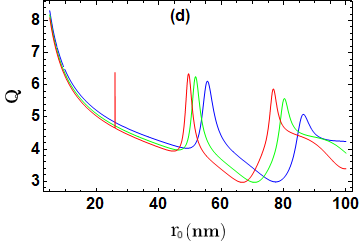}\\
	\includegraphics[scale=0.34]{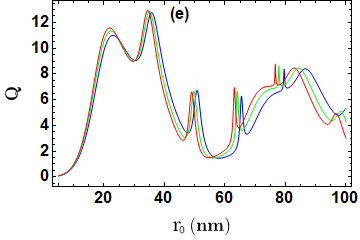}\includegraphics[scale=0.34]{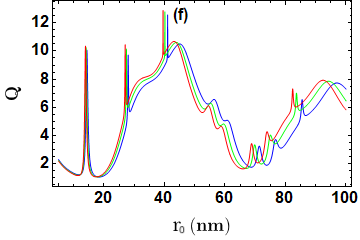}\includegraphics[scale=0.34]{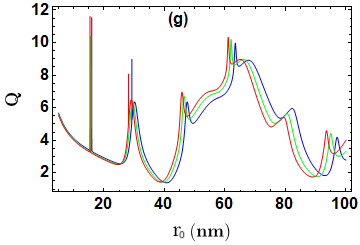}\includegraphics[scale=0.34]{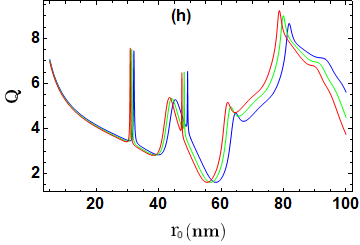}
	\caption{ The scattering efficiency $Q$ as a function of of the radius $r_0$ for $E = 20 \text{ meV}$ and $\mathcal{P} = 0.3$ by changing the values of the value of electrostatic potential $U$ such that blue line for $U = 0 \text{ meV}$, green line for $U = 5 \text{ meV}$ and red line for $U = 10 \text{ meV}$. In the ours plots, we plot by varying magnetic flux $\Phi$ such that (a, e) for $\Phi = 0$, (b, f) for $\Phi = 1/2$, (c, g) for $\Phi = 3/2$ and (d, h) for $\Phi = 5/2$. We plot by fixing the value of light intensity $I_L$ such that (a, b, c, d) for  $I_L = 3 \text{ W/cm}^2$ and (e, f, g, h) for  $I_L = 5 \text{ W/cm}^2$.}
	\label{fig4}
\end{figure}

As shown in Fig. \ref{fig5}, a comprehensive study of the scattering efficiency, denoted $Q$, as a function of the incident electron energy $E$, is presented for a graphene quantum dot with radius $r_0 = 70 \text{ nm}$ and light intensity $I_L = 3.5 \text{ W/cm}^2$. The present study is divided into eight panels, which explore the interplay between the light polarization $\mathcal{P}$, the magnetic flux $\Phi$, and the electrostatic potential $U$. 
As can be seen in Fig. \ref{fig5}, there is a strong dependence of the scattering efficiency $Q$ on the incident energy $E$, marked by the appearance of resonance peaks at specific energies. These peaks are not features, but rather undergo a shift to higher energies according to the value of the applied magnetic flux. This observation implies a close correlation between the magnetic confinement and the energy levels accessible to the electrons. The resonance peaks show a transformation in both amplitude and configuration, which are found to be {dependent} on the energy levels, the electrostatic potential $U$, as well as the magnetic flux. In particular, we observe that the number of resonances decreases with increasing flux. The observation of these resonances underscores the non-uniform nature of electron scattering as a function of incident energy and could be related to the excitation of quasi-bound states within the GQDs and how the electron energy alters the possibility of excitation of these states, as highlighted in studies of electron trapping in quantum dots (see \cite{Penalight,azar24}). 
The influence of the light polarization $\mathcal{P}$ is also evident, as shown by the curves associated with $\mathcal{P} = 0.3$, $0.4$, and $0.5$. Despite the minimal differences between these curves, the subtle variations in the energy peaks and their amplitudes suggest that the polarization of light can modulate the way electrons are scattered by interaction with light. This finding also suggests that light can modulate the intensity of the electron-light interaction, a phenomenon that could be related to studies of the effect of quasi-bound states as a function of light polarization \cite{Penalight}.
An important effect is the interaction of the magnetic flux $\Phi$ and the electrostatic potential $U$, which modifies the structure of the scattering resonances, influencing their amplitudes, number and energy positions. It has been observed that a high value of $U$ leads to a reduction in the amplitude of the peaks and a suppression of the resonances. These results suggest a correlated influence of electrostatic and magnetic parameters on the electron scattering behavior.
   \begin{figure}[ht]
	\centering
	\includegraphics[scale=0.34]{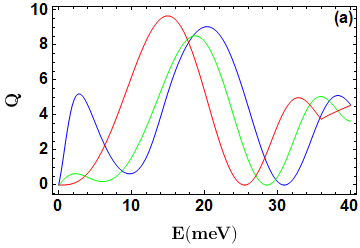}\includegraphics[scale=0.34]{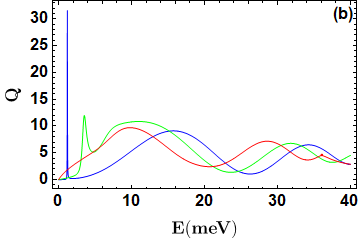}\includegraphics[scale=0.34]{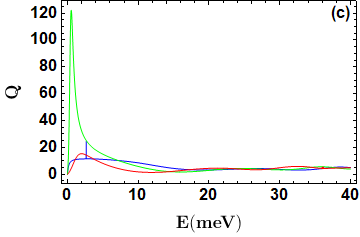}\includegraphics[scale=0.34]{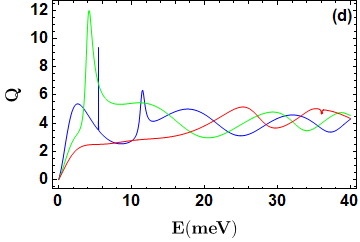}\\
	\includegraphics[scale=0.34]{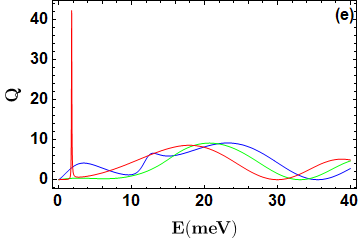}\includegraphics[scale=0.34]{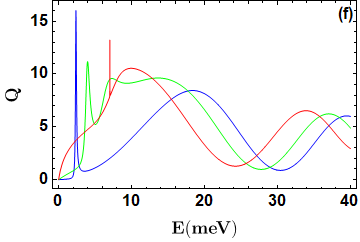}\includegraphics[scale=0.34]{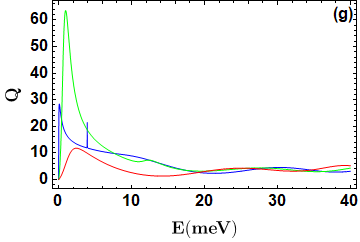}\includegraphics[scale=0.34]{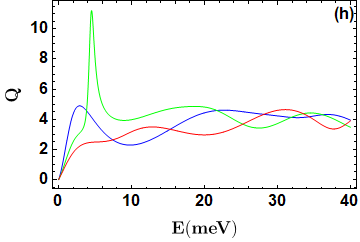}
	\caption{ The scattering efficiency $Q$ as a function of the incident energy $E$ for $r_0 = 70 \text{ nm}$ and $I_L = 3.5 \text{ W/cm}^2$ by changing the value of light polarization $\mathcal{P}$ such that blue line for $\mathcal{P} = 0.3$, green line for $\mathcal{P} = 0.4$ and red line for $\mathcal{P} = 0.5$. In our plots, we plot by varying magnetic flux $\Phi$ such that (a, e) for $\Phi = 0$, (b, f) for $\Phi = 1/2$, (c, g) for $\Phi = 3/2$ and (d, h) for $\Phi = 5/2$. {The lots were generated} by fixing the value of electrostatic potential $U$ such that (a, b, c, d) for $U = 0 \text{ meV}$ and (e, f, g, h) for $U = 10 \text{ meV}$.}
	\label{fig5}
\end{figure}

As shown in Fig. \ref{fig6}, the influence of the rescaled magnetic flux $\Phi$ on the scattering efficiency $Q$ is studied in detail in a system where the incident energy $E$ is fixed at $20 \text{ meV}$ and the radius $r_0$ is $70 \text{ nm}$. The analysis reveals a complex relationship characterized by significant oscillations of $Q$ as a function of $\Phi$, indicating resonance phenomena where scattering becomes particularly effective. These oscillations, which appear as pronounced peaks, vary not only in amplitude but also in frequency and shape as system conditions are changed.
The first observations show that the amplitude of these oscillations {increases} when the magnetic flux values are reduced. This finding indicates that the interaction of electrons with light is more effective in circumstances where the magnetic flux is less pronounced. Conversely, as the magnetic flux increases, the amplitude of these oscillations decreases, which means a decrease in the efficiency of scattering. This phenomenon is particularly evident when the light intensity of $I_L$ is low, as can be seen by comparing Figs. \ref{fig6}\bb{(a)-(d)} and Figs. \ref{fig6}\bb{(e)-(h)}. This result underscores the importance of light intensity in shaping the nature of electron-quantum dot interactions, as a higher value of $I_L$ enhances the visibility of amplitude oscillations in $Q$.
As shown in \ref{fig6}\bb{(a)-(d)}, the electrostatic potential $U$ significantly affects the diffusion efficiency. A comparison of the different curves within each panel shows that the amplitude of the resonance peaks decreases as $U$ increases, and the peaks become less distinct. This phenomenon is consistent with the observations in Fig. \ref{fig4}, which showed that scattering decreases as $U$ increases. This observation suggests that the electrostatic potential has a {destructive} effect on the scattering process by affecting the ability of electrons to penetrate the quantum dot and thereby create scattering resonances. Fig. \ref{fig6} reveals another peculiarity: unlike Fig. \ref{fig4} and Fig. \ref{fig5}, light plays a key role in the $Q$ oscillations, and these are highly polarization dependent, demonstrating the sensitivity of the electrons to the polarization state of the light. For example, the peak density is higher and more dispersed for a polarization value of $0.4$, and lower and sharper for a value of $0.3$.
In terms of the combined behavior exhibited, it is evident that the parameters under study {are correlated and} interact in complex ways to determine the efficiency of scattering. The influence of magnetic flux and electrostatic potential on resonant structures is evident, while light polarization has been shown to modulate amplitudes. It is evident that the intensity of the light has a direct correlation with the amplitude and efficiency of scattering, with higher light intensities resulting in larger values of these parameters. This observation suggests that a thorough understanding of the scattering mechanisms in this system requires an approach that considers all of these parameters and their interactions in a unified framework.

\begin{figure}[ht]
	\centering
	\includegraphics[scale=0.34]{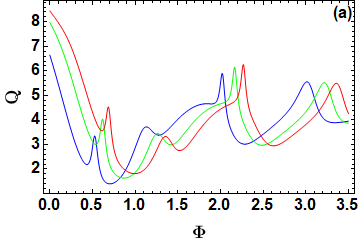}\includegraphics[scale=0.34]{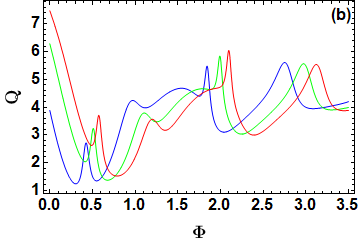}\includegraphics[scale=0.34]{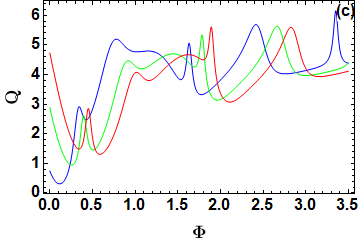}\includegraphics[scale=0.34]{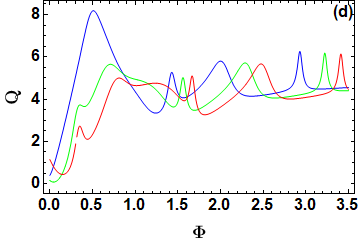}\\
	\includegraphics[scale=0.34]{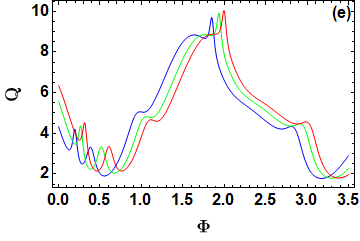}\includegraphics[scale=0.34]{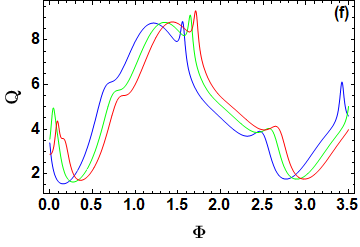}\includegraphics[scale=0.34]{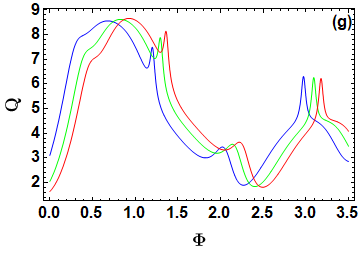}\includegraphics[scale=0.34]{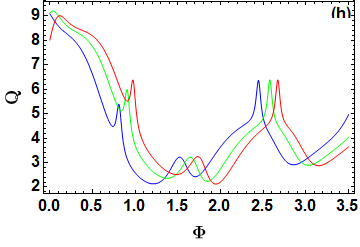}
	\caption{ The scattering efficiency $Q$ as a function of the rescaled magnetic flux $\Phi$ for $E = 20 \text{ meV}$ and $r_0 = 70 \text{ nm}$ by changing the values of the electrostatic potential $U$ such that blue line for $U = 0 \text{ meV}$, green line for $U = 5 \text{ meV}$ and red line for $U = 10 \text{ meV}$. In our plots, we plot by varying light polarization $\mathcal{P}$ such that (a, e) for $\mathcal{P} = 0.3$, (b, f) for $\mathcal{P} = 0.4$, (c, g) for $\mathcal{P} = 0.5$ and (d, h) for $\mathcal{P} = 0.6$. We plot by fixing the value of light intensity $I_L$ such that (a, b, c, d) for $I_L = 3 \text{ W/cm}^2$ and (e, f, g, h) for $I_L = 5 \text{ W/cm}^2$.}
	\label{fig6}
\end{figure}

 Fig. \ref{fig8} presents a comprehensive analysis of the spatial distribution of electrons {density plot} in a GQD, elucidating the complex behavior under varying conditions. It is important to note that all plots were made while keeping the basic parameters constant. The radius of the quantum dot is fixed at $r_0=70 \text{ nm}$, the energy of the incident electrons at $E-20 \text{ meV}$, and the light intensity at $3 \text{ W/cm}^2$. To gain a deeper understanding of the mechanisms of electron localization, we have systematically investigated the influence of several key parameters. The study is based on two different scattering modes ($m=0$ and $m=1$), each investigated under different electrostatic potential conditions. For each mode, the initial condition was set with no electrostatic potential ($U=0 \text{ meV}$), followed by a potential of $10 \text{ meV}$. In each configuration, the magnetic flux was gradually varied, with three specific values explored: $0$, $1/2$, and $3/2$, respectively. This methodical approach, illustrated by twelve different configurations (panels \bb{(a)} to \bb{(l)}), allows to highlight the combined influence of electrostatic potential, magnetic flux and light polarization on the confinement properties of electrons in the GQD. The results obtained for the $m=0$ mode are shown in Figs. \ref{fig8}\bb{(a)-(f)}, while those for the $m=1$ mode are shown in Figs. \ref{fig8}\bb{(g)-(l)}.
As shown in Fig. \ref{fig8}\bb{(a)-(c)} and Fig. \ref{fig8}\bb{(g)-(i)}, the total scattering efficiency is high when the electrostatic potential is zero. Furthermore, it is generally observed that the electron density distribution is sensitive to the scattering mode. For $m=0$ (Figs. \ref{fig8}\bb{(a)-(c))}, the density is more {concentrated} outside the GQDs and forms radial patterns with spatial modulations indicative of wave interaction. Conversely, as the magnetic flux increases, the patterns become less pronounced and the density becomes more homogeneous. Conversely, for $m=1$ (Figs. \ref{fig8}\bb{(g)-(i))}), the density is more localized at the edge of the GQDs, {signaling the dominance of edge modes}, and the interference patterns are more pronounced. It is evident that the localization at the edge is more pronounced for a flux of $\Phi=0$ and becomes more homogeneous as the magnetic flux increases. 
A comparison of Fig. \ref{fig8}\bb{(a)} and Fig. \ref{fig8}\bb{(b)}, for a flux of either $0$ or $1/2$, reveals a variation in the density of the $m=0$ mode, manifested as structures extending in the x and y directions. Conversely, a similar structure is observed for the $m=1$ mode, but with {increased} density amplitudes at the periphery of the GQD. The effect of the magnetic flux is evident in Figs. \ref{fig8}\bb{(b,c)}, where the interference structure is less pronounced, {while in} \ref{fig8}\bb{(h,i)}, the density is more dispersed around the GQDs. This observation indicates that an increase in magnetic flux leads to a change in the direction of electron scattering, confirming the significant role of magnetic confinement in determining the density distribution.
As shown in Figs. \ref{fig8}\bb{(d)-(f)} and Fig. \ref{fig8}\bb{(j)-(l)}, it is {clear} that the application of an electrostatic potential of $U = 10 \text{ meV}$ significantly affects the spatial distribution of the density. In general, the probability density is more homogeneous and less intense, with weaker localization at the edge of the quantum dot. A comparison of the different panels for $U = 10 \text{ meV}$ shows that the polarization dependence is weaker than for $U = 0$.
A comparison of Figs. \ref{fig8}\bb{(d,e)}, corresponding to the $m=0$ mode, with Figs. \ref{fig8}\bb{(j,k)}, corresponding to the $m=1$ mode, shows that the scattering in the $m=0$ mode is more spatially dispersed and has a weaker intensity compared to the $m=1$ mode.
Furthermore, the analysis of Figs. \ref{fig8}\bb{(e,f)} and \ref{fig8}\bb{(k,l)} shows that the increase of the magnetic flux leads to a more uniform distribution of the electron density and a reduction of its amplitude, a phenomenon consistent with the observations presented in Fig. \ref{fig4}, where the resonances appear less pronounced for higher flux values.
As shown in Fig. \ref{fig8}, the density distributions can be correlated with the results obtained for the diffusion efficiency $Q$ in Fig. \ref{fig4} and Fig. \ref{fig5}. The resonances of $Q$ are more pronounced when the density is highly localized at the edge of the GQD and less intense when the density is more diffused in space. For example, the results presented in this figure show that a zero or low magnetic flux and a zero or low electrostatic potential are conducive to a pronounced localization of the electrons at the edge of the quantum dot, a condition that leads to intense scattering resonances. This edge localization of GQDs is consistent with the results reported in \cite{Penalight}, which showed that edge localization exists for quasi-bound states.
In conclusion, Fig. \ref{fig8} shows that the scattering efficiency discussed in Fig. \ref{fig4} and Fig. \ref{fig5} is closely related to the spatial distribution of the electron density in the vicinity of the GQDs. The polarization of the light and the magnetic flux are found to play a crucial role in the configuration of the density patterns and their amplitude. The results thus obtained demonstrate how the interplay of these parameters governs the propagation and confinement of electrons inside {and outside} the GQDs.
\begin{figure}[H]
	\centering
	\includegraphics[scale=0.46]{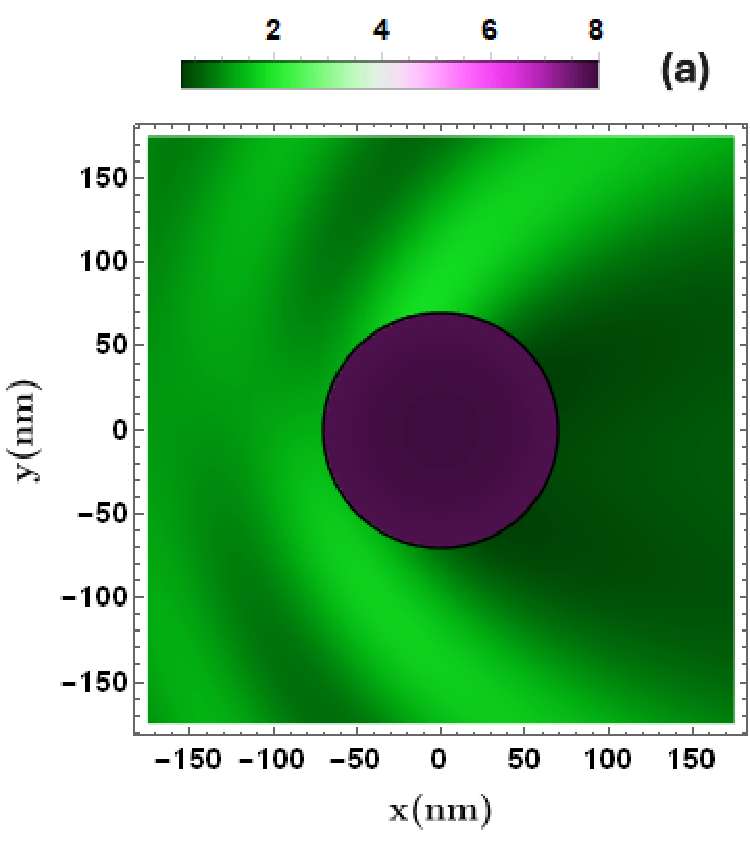}\includegraphics[scale=0.46]{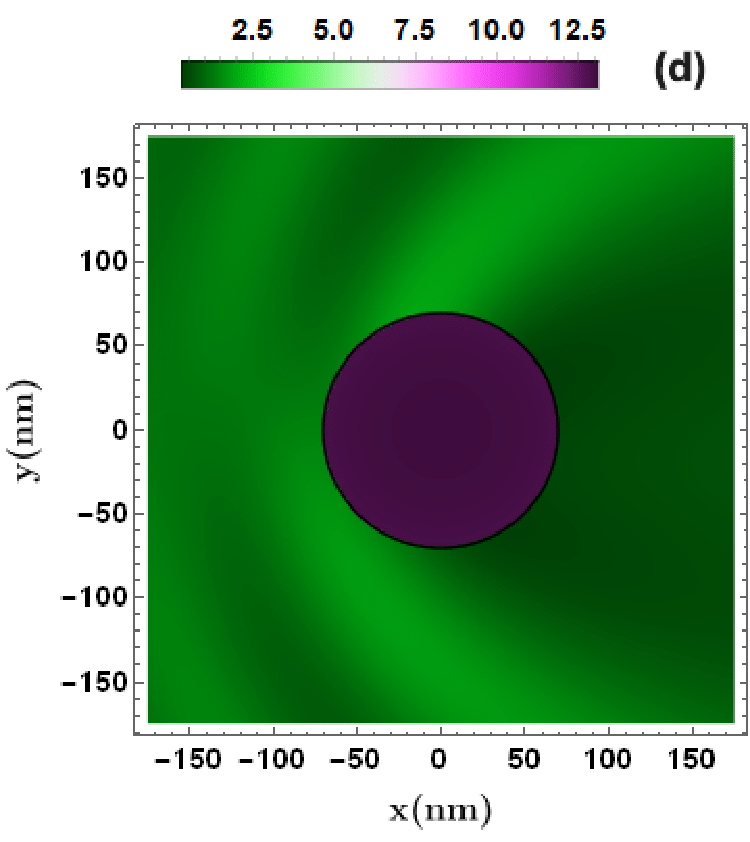}\includegraphics[scale=0.46]{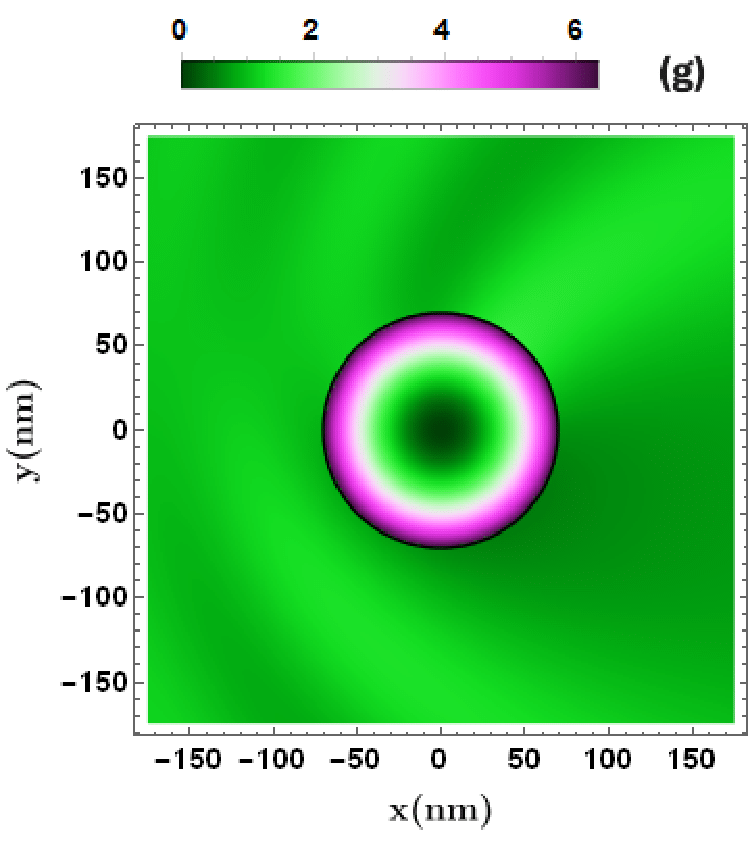}\includegraphics[scale=0.46]{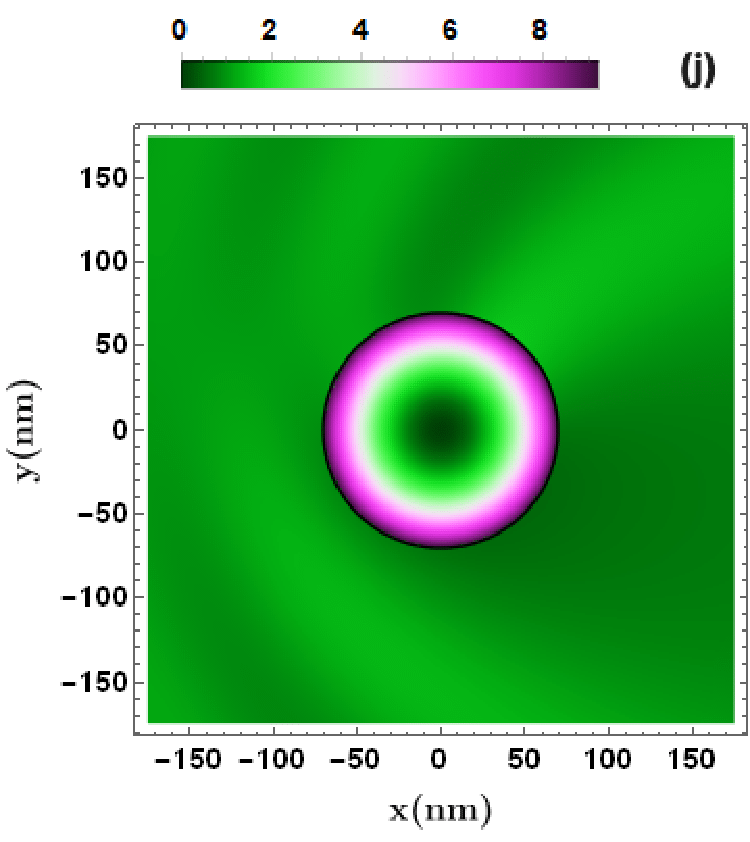}\\
	\includegraphics[scale=0.46]{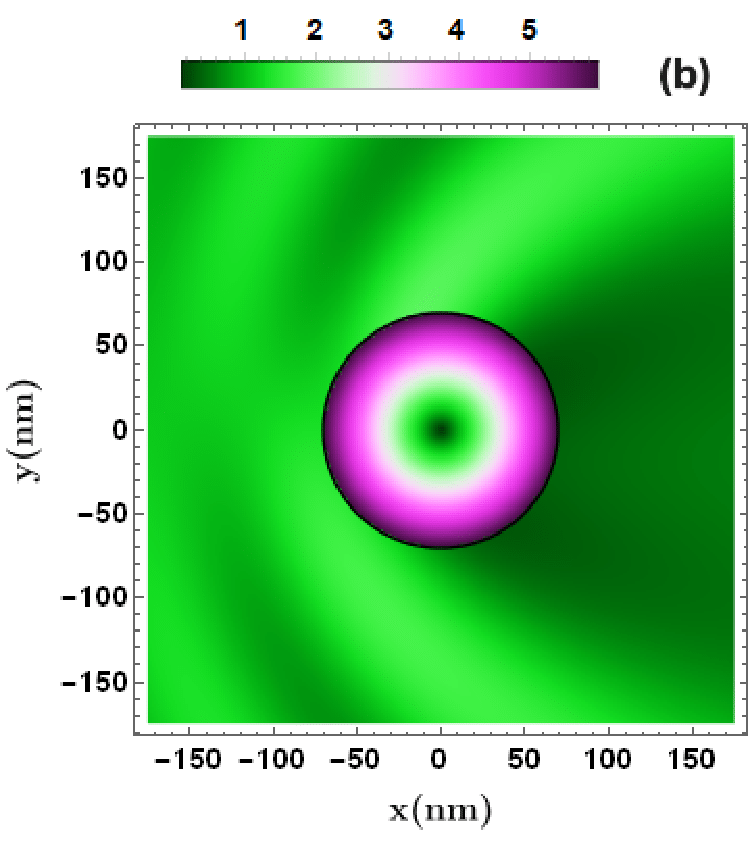}\includegraphics[scale=0.46]{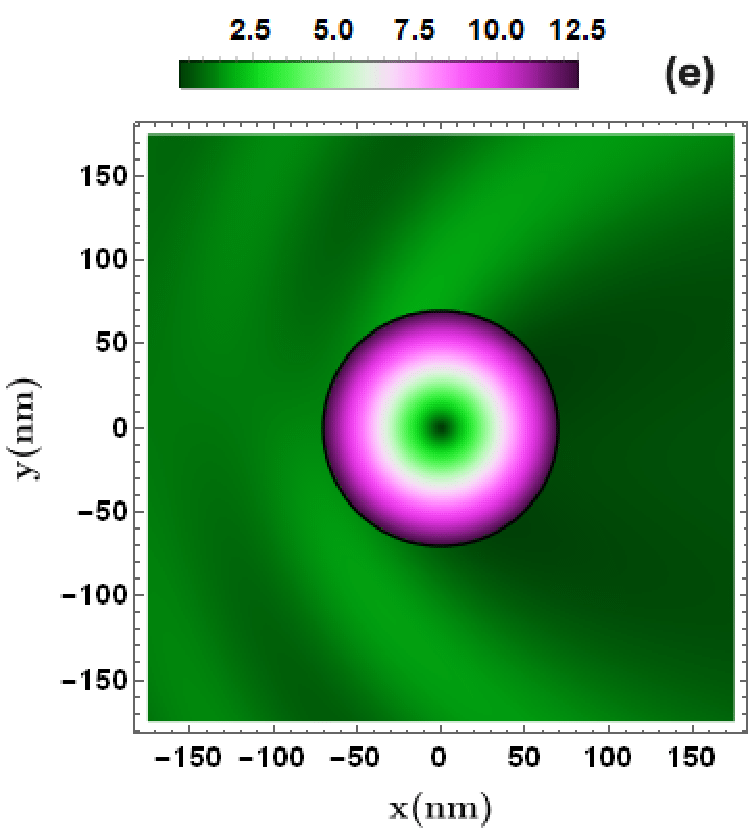}\includegraphics[scale=0.46]{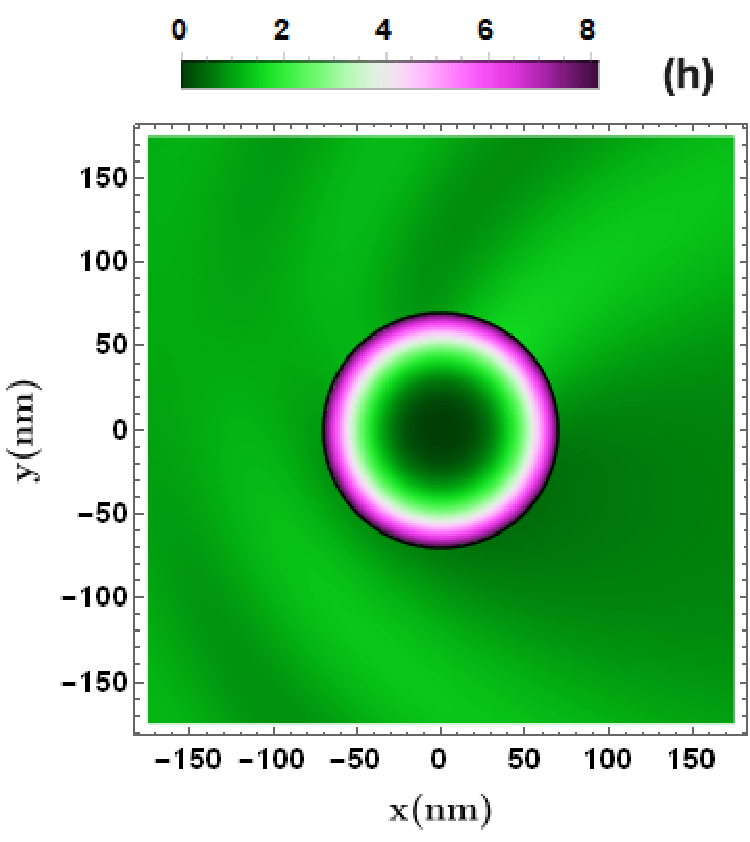}\includegraphics[scale=0.46]{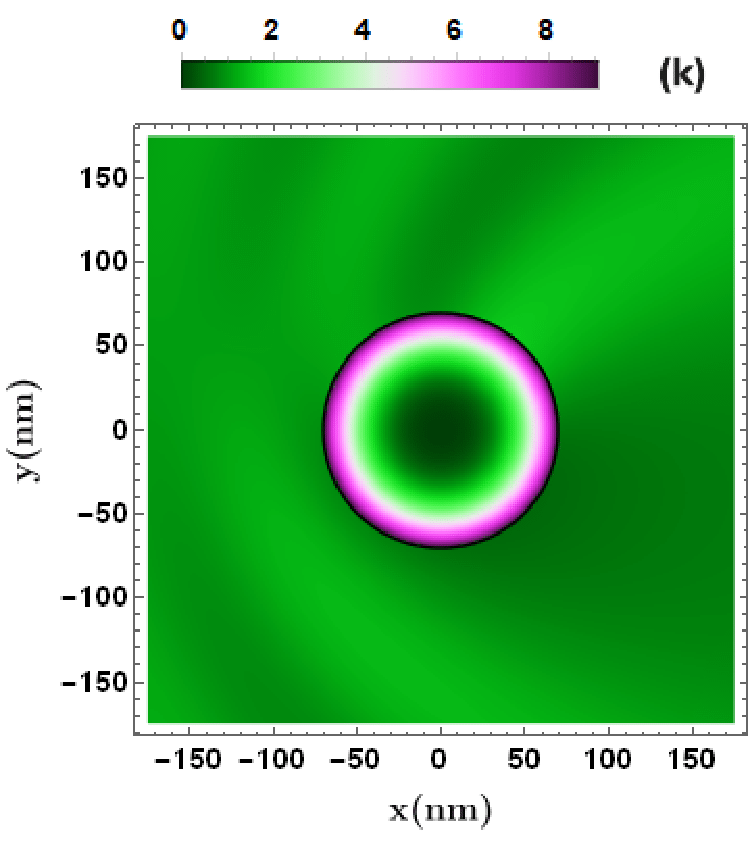}\\ 
	\includegraphics[scale=0.46]{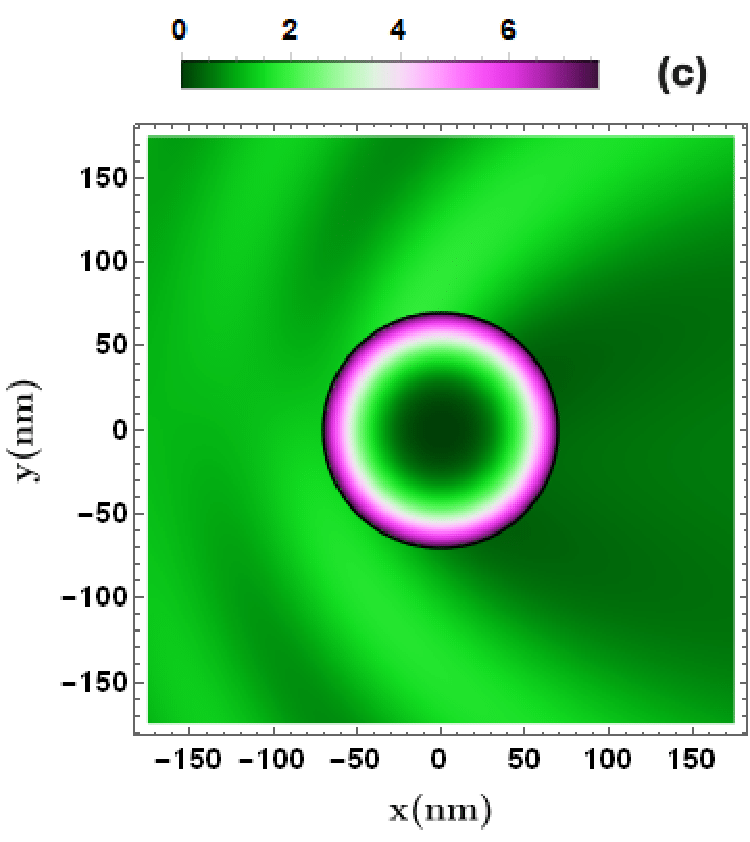}\includegraphics[scale=0.46]{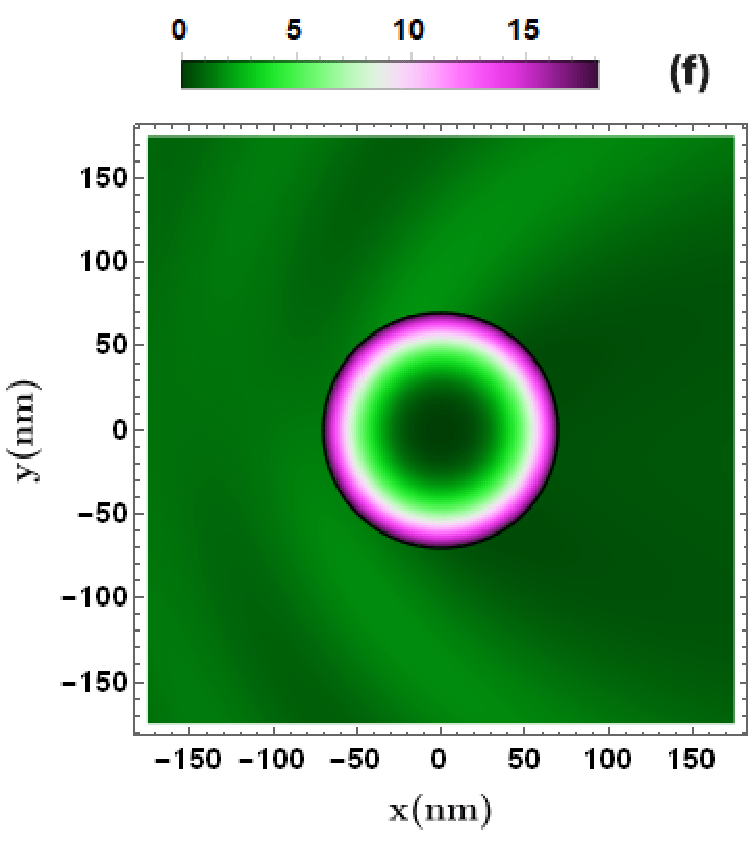}\includegraphics[scale=0.46]{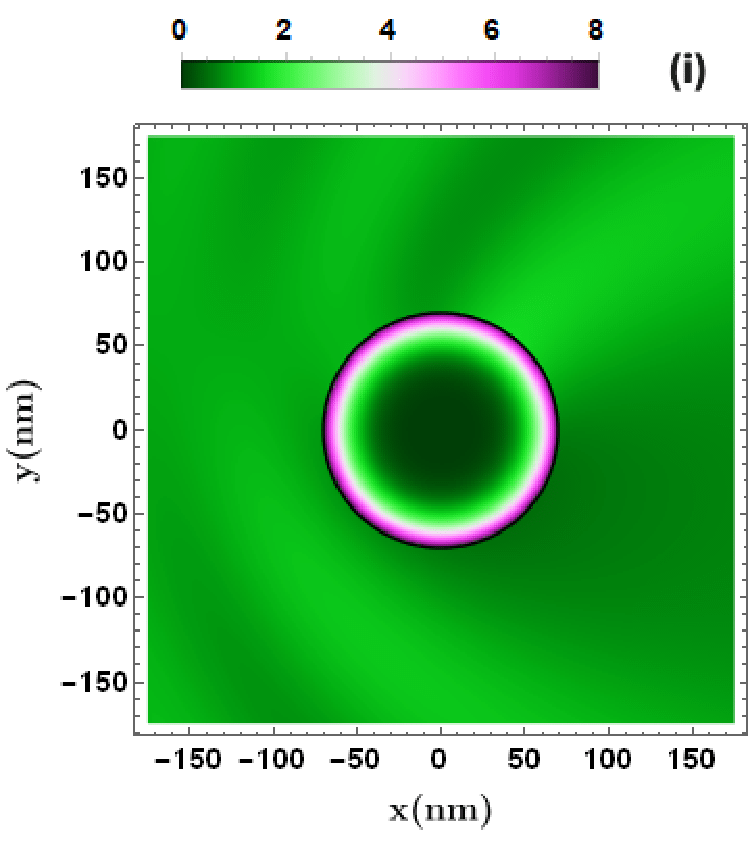}\includegraphics[scale=0.46]{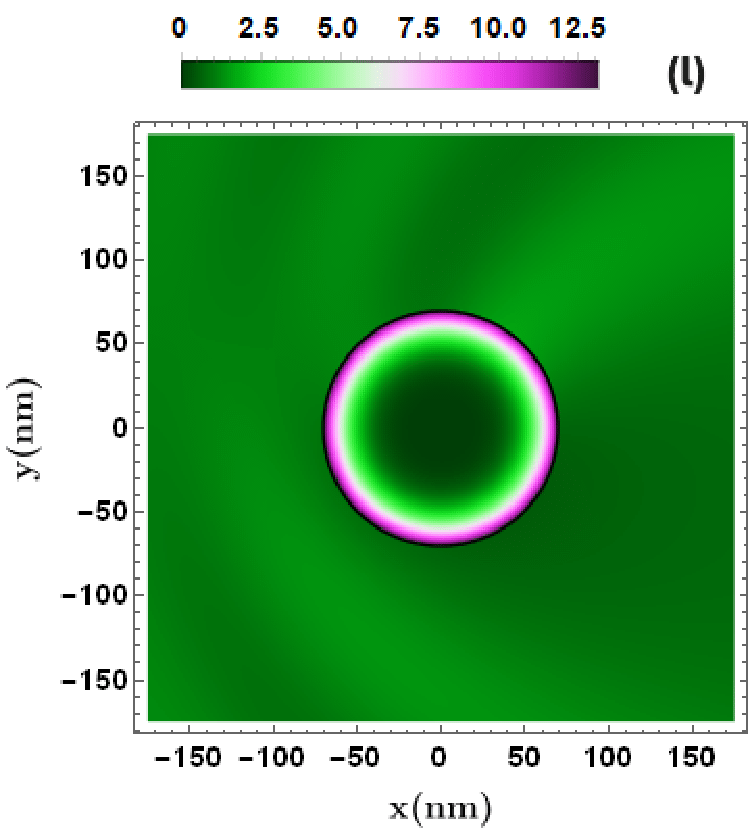}
	\caption{ Scattering analysis in terms of density for a real space examination of electron scattering on a GQD for $E = 20 \text{ meV}$, $I_L = 3 \text{ W/cm}^2$ and $r_0 = 70 \text{ nm}$ by changing the values of parameters ($m$, $\mathcal{P}$, $U$, $\Phi$) such that (a, b, c, d, e, f) for $m= 0$ and (g, h, i, j, k, l) for $m =1$ but the value of $\mathcal{P}$ is fixed where (a) for $\mathcal{P} = 0.714$, (b) for $\mathcal{P} = 0.713$, (c) for $\mathcal{P} = 0.713$, (d) for $\mathcal{P} = 0.867$, (e) for $\mathcal{P} = 0.867$, (f) for $\mathcal{P} = 0.868$, (g) for $\mathcal{P} = 0.72$, (h) for $\mathcal{P} = 0.715$, (i) for $\mathcal{P} = 0.715$, (j) for $\mathcal{P} = 0.867$, (k) for $\mathcal{P} = 0.867$, and (l) for $\mathcal{P} = 0.868$. In ours plots, we plot by fixing the value of $U$ such that (a, b, c, g, h, i) for $U =0 \text{ meV}$ and (d, e, f, j, k l) for $U = 10 \text{ meV}$. The magnetic flux $\Phi$ is varied as follows: (a, d, g, j) correspond to $\Phi = 0$, (b, e, h, k) correspond to $\Phi = 1/2$, and (c, f, i, l) correspond to $\Phi = 3/2$.}
	\label{fig8}
\end{figure}

\section{Conclusion}\label{conc}

Our theoretical study presents a thorough and systematic analysis of the interaction between Dirac electrons and a graphene quantum dot (GQD) simultaneously exposed to a magnetic flux and circularly polarized light irradiation. The theoretical model developed here is based on the Dirac-Weyl equation and it elucidates how the electronic states, both inside and outside the GQDs, are affected by the collective action of these two external parameter fields.
The numerical analysis, which is the cornerstone of our study, shows that light polarization and intensity play a crucial role in modulating the electron scattering. In particular, light polarization exerts a significant influence on the electron localization in the vicinity of the GQD. The results also show that the spatial distribution of the probability density, an indicator of the quantum nature of electron transport, is strongly influenced by these external parameters. In particular, variations in light intensity allow precise manipulation of the scattering process, which can be linked to the influence of light on the formation of quasi-bound states. Furthermore, by studying the contributions of the different scattering modes, we were able to clarify how the electrons interact with the quantum dot, and the reasons why these interactions are more or less influential depending on the energy, flux, and potential.

It has been shown that magnetic flux and electrostatic potential affect the scattering mechanisms by modifying the amplitude and shape of the resonances, thereby underscoring the complexity of the interaction of electrons with their electromagnetic environment. It is important to note that these parameters are not independent; rather, they collectively influence the spatial distribution of electrons. The density distribution is generally more intense when the electrostatic potential is zero, {in} the presence of magnetic flux, although reducing the intensity, gives rise to density reshaping with interference-related scattering patterns.
Furthermore, the results obtained demonstrate the possibility of controlling the lifetime of electron quasi-bound states within the GQD via the intensity and polarization of light. These quasi-bound states, which correspond to scattering resonances, can be precisely manipulated by changing the intensity or polarization of light, and this mechanism can also be achieved via magnetic flux. The exploitation of these interactions has the potential to open new avenues for the design of innovative charge carriers in graphene-based nanostructures.

 The demonstration of the possibility to finely control the lifetime of electron quasi-bound states via light intensity and polarization is a significant finding with promising potential applications in optoelectronic sensors, biosensors, and quantum information technology. The study of other types of structured light, such as light polarized in a vector mode or with orbital angular momentum, is a promising avenue for future investigations. Experimental validation of our theoretical predictions would be a crucial step towards the realization of innovative graphene-based quantum devices.

\end{document}